\documentstyle[aas2pp4]{article}

\slugcomment{\it Submitted to the Astrophysical Journal Letters}

\lefthead{Simard and Pritchet}
\righthead{Kinematics of Intermediate Redshift Galaxies}

\begin{document}

\title{Internal Kinematics of Galaxies at z = 0.25$-$0.45}

\author{Luc Simard\altaffilmark{1} and Christopher J. Pritchet\altaffilmark{1}}
\affil{Department of Physics and Astronomy, University of Victoria, P.O. Box 
3055, Victoria, Canada, V8W 3P6, E-mail: 
(simard, pritchet)@uvastro.phys.uvic.ca}


\altaffiltext{1}{Visiting Astronomer, Canada-France-Hawaii Telescope, which 
is operated by the National Research Council of Canada, le Centre National 
de la Recherche Scientifique of France, and the University of Hawaii.}


\vskip 24 true pt

\centerline{\it Submitted to Astrophysical Journal Letters}

\begin{abstract}

Low-mass starbursting galaxies have been proposed as the 
explanation of the excess of faint galaxies observed at 
intermediate redshifts. If this hypothesis is correct, then 
intermediate redshift galaxies should rotate more 
slowly than nearby galaxies with the same rest-frame luminosity. 
We present the results of a survey of the internal kinematics of 
intermediate redshift (z = 0.25$-$0.45) field galaxies to search 
for this effect. Using the Canada-France-Hawaii Telescope, 
spatially-resolved spectra of the [O II] $\lambda\lambda$ 
3726$-$3729 \AA
\thinspace\thinspace doublet emission line have been obtained 
for 22 galaxies. V$_{\rm rot}$ sin $i$ and [O II] disk scale lengths 
have been extracted from each galaxy spectrum using a Bayesian fitting 
technique.

Galaxies in the sample are found to be $\sim$1.5$-$2.0 mag brighter 
than expected from their rotation velocity and the local Tully-Fisher 
(TF) relation. Low-mass galaxies exhibit a wider range of evolution 
relative to the TF relation than high-mass galaxies. The main source of 
uncertainty in this result is the large scatter in the local TF 
relation for late-type galaxies. Luminosity-dependent luminosity 
evolution neatly reconciles the lack of evolution seen in other works 
with the results of our survey. It is also found that the overall 
properties of [OII] kinematics at intermediate redshifts are varied. 
For example, 25$\%$ of the field galaxies in the sample have [OII] 
kinematics unrelated to rotation; [OII] emission is confined to the 
nucleus in most of these galaxies. Anomalous kinematics is found to 
be related to the presence of companions --- i.e. minor merger events. 
A Doppler ellipse similar to those found in local dwarf galaxies has 
been observed in a z = 0.35 galaxy, and may be interpreted as a 
supernova-driven supershell. 

\end{abstract}




%

\section{Introduction}

The observed number counts of faint galaxies exceed the predictions of 
standard models in which evolutionary effects are ignored. Yet the faint 
galaxy redshift distribution appears to be well modelled by these same 
no-evolution models (Lilly {\it et al.} 1995, Lilly 1993, Koo and Kron 
1992, and references therein). The amount of evolution that faint
galaxies have undergone (relative to galaxies at the present epoch) is
clearly a critical parameter for understanding the Universe at 
intermediate redshifts. Here we focus on the novel technique of 
internal kinematics for studying galaxy evolution.

Internal kinematics is directly related to a single fundamental property 
of galaxies: mass. Observations of the velocity field of intermediate 
redshift galaxies can therefore be used to quantify the amount of 
luminosity evolution that these galaxies have undergone. As an 
illustration, consider a galaxy with rest-frame absolute luminosity 
equal to that of the Milky Way. If this galaxy were also as massive 
as the Milky Way (i.e. if it were unevolved in luminosity relative to 
the Milky Way), then a rotation velocity of $\sim$ 200 km/s would be 
observed. On the other hand, if it were in fact a lower mass object 
that had been boosted in luminosity by $\sim$10$\times$ (as suggested 
in the luminosity-dependent luminosity evolution scenario -- 
Broadhurst {\it et al.} 1988), then the TF relation would predict 
a rotation velocity $\sim$ 100 km/s. 

This kinematical approach to measuring evolution is direct. It is not 
affected by uncertainties in models based on local luminosity functions. 
Furthermore, whereas luminosity functions derived from redshift surveys 
constrain the average amount of evolution in an entire galaxy population, 
the present approach can measure luminosity enhancement in individual 
galaxies, and can thus tie luminosity evolution directly to galaxy 
properties on a {\it galaxy by galaxy} basis. 

Most previous kinematical studies (Franx 1993, Vogt {\it et al.} 1993, 
Colless 1994, Koo {\it et al.} 1995, Forbes {\it et al.} 1996) have 
suffered either from small sample size, or from a lack of spatial 
information; but they have clearly demonstrated that current telescopes 
are adequate to the challenge of measuring the internal kinematics of 
intermediate redshift galaxies. Recently, Vogt {\it et al.} (1996) 
presented a beautiful set of rotation curves (observed with the Keck 
Telescope) for nine faint field galaxies in the redshift range 0.1 
$\leq$ z $\leq$ 1. These rotation curves appeared similar to those 
of local galaxies in both form and amplitude. The kinematics of the 
Keck galaxies showed evidence for only a modest increase in luminosity 
($\Delta$M$_B$ $\leq$ 0.6 mag) compared to the local Tully-Fisher 
relation. This is in apparent disagreement with the strong evolution 
in galactic disks observed by Schade {\it et al.} (1996) over the range 
0.1 $<$ z $<$ 0.6. 

In this paper, we present the internal kinematics of a sample of 
intermediate redshift galaxies, and demonstrate how luminosity-dependent 
luminosity evolution can reconcile observations of surface brightness 
and internal kinematics. 

\section{Observations}

Twenty field galaxies and two cluster galaxies in the redshift range 
0.25$-$0.45 were selected from the CNOC cluster survey database (Yee 
{\it et al.} 1996). This database contains position, Gunn $r$ magnitude, 
$g-r$ color, and redshift for thousands of galaxies. Elongated objects 
were preferentially selected to minimize sin $i$ effects. No size/color 
selection criteria were used. Rest-frame [OII] equivalent widths W$_{3727}$ 
were chosen to lie between 20 and 50 \AA. Target galaxies may be [OII] 
strong relative to local early-type (Sb and earlier) spirals, but they 
have the same W$_{3727}$ as many local late-type (Sc and later) galaxies 
(Kennicutt 1992). Our range of W$_{3727}$ is also representative of the 
excess galaxy population at intermediate redshifts (Broadhurst {\it et 
al.} 1988, Broadhurst {\it et al.} 1992). 

Spatially-resolved observations of the [O II] $\lambda$$\lambda$ 3726.1, 
3728.8 \AA\thinspace\thinspace doublet emission line were obtained with 
the Multi-Object 
Spectrograph (MOS) and the Subarcsecond Imaging Spectrograph (SIS) 
at the 
Ca-nada-France-Hawaii 3.6-m Telescope (CFHT) in July-August 1994. 
SIS was ideally suited for studying [OII] kinematics because SIS ``tip-tilt'' 
corrections yielded a seeing FWHM of 0\arcsec.5$-$0\arcsec.6. We 
used the B600 grism and a LORAL 2048$\times$2048 CCD (QE = 23$\%$, N$_R$ 
= 8 e$^{-}$). The detector was binned by 2 along both axes to improve S/N 
ratio. Image scale and dispersion were 0\arcsec.31/pixel and 1.58 \AA/pixel 
(MOS) and 0\arcsec.17/pixel and 0.88 \AA/pixel (SIS). Typical total 
integration time per galaxy was 7200 seconds split in 2$-$3 exposures for 
cosmic ray removal. 





\section{Synthetic Rotation Curve Fitting}

Flux levels in [O II] were very low (typical S/N ratio {\it per pixel} 
$\sim$ 2$-$3); it was therefore important to choose a method of analysis 
which used all of the pixels simultaneously to statistically find the best 
parameter values and their respective uncertainties. We adopted a synthetic 
rotation curve fitting technique, in which models were compared to the two 
dimensional distribution of pixel intensities in the [O II] line (the two 
dimensions being wavelength or velocity, and position along the slit). The 
parameters of the fitting model were the projected rotation velocity V$_
{\rm rot}$ sin $i$ of the galaxian disk in km/s, the [OII] exponential disk 
scale length r$_d$ in h$^{-1}$ kpc, the [OII] total line flux in DU, and 
the dimensionless [OII] doublet intensity ratio I(3726 \AA)/I(3729 \AA). 
(It was necessary to include this doublet intensity ratio in the model 
because the [O II] doublet is partially resolved due to (1+z) spectral 
stretching.)

The 2D model distributions of pixel intensities were constructed assuming
that the [OII] emission was distributed in a thin exponential disk with a flat 
rotation curve. These [OII] model disks were convolved with a 
point-spread-function (PSF) extracted from direct images; the effect of 
placing a slit in front of the image was then computed. Finally, the flux 
passing through 
the synthetic slit was convolved with the spectrograph instrumental line 
shape, which was extracted from comparison arc lines. 

The best fitting parameter values were found using the Metropolis algorithm 
(Saha and Williams 1994). This Bayesian algorithm starts with some set of 
parameter values $\omega$ and the associated $P(\omega|D,M)$ --- the 
probability that $\omega$ is the true parameter set given the data D and 
a model M. It then picks a possible change $\delta \omega$ in the parameters 
and computes $P(\omega + \delta \omega |D,M)$. If $P(\omega + \delta \omega 
|D,M) > P(\omega|D,M)$, then the change $\delta \omega$ is accepted. If 
$P(\omega + \delta \omega |D,M) <  P(\omega|D,M)$, then the change $\delta 
\omega$ is accepted only some fraction $P(\omega + \delta \omega |D,M)$/
$P(\omega|D,M)$ of the time. After hundreds of iterations, the distribution 
of accepted $\omega$ values will converge to $P(\omega|D,M)$ provided that 
all possible $\omega$ are eventually accessible. Parameter space is thus 
sampled with a density proportional to the likehood. 

The trial changes $\delta \omega$ are chosen at random. If they are too 
small (i.e. the parameter search is too ``cold''), then all iterations 
will accept changes. If the search is too ``hot'', then none of the 
iterations will accept changes. The ``temperature'' of the search is 
regulated so that half the iterations accept changes. As the algorithm 
proceeds, trial values are stored to simultaneously derive the best 
parameter values and their respective error distribution. For each 
parameter, we took the median value of $P(\omega|D,M)$ as the best value 
and the 16th and 84th percentile values as our ``1\,$\sigma$'' error 
bars. We tested confidence intervals and parameter value recovery using 
a set of 342 simulated rotation curves. We chose input parameter values 
in these simulations to yield worse S/N ratios than seen in the data. 
No biases were detected. The Metropolis parameter estimates were also 
more robust to noise than least-squares estimates at low S/N ratios.


\section{Results}

\subsection{[OII] Morphologies} \label{o2morph}

[OII] kinematics at intermediate redshifts is varied. Seven galaxies 
have [O II] scale lengths significantly smaller than their broad-band 
scale lengths, consistent with central unresolved [O II] sources. Two 
of these ``unresolved [O II]'' galaxies are serendipitously observed 
cluster galaxies (we cut extra slitlets on the mask for cluster galaxies 
whenever possible). Such discrepancies indicate that the [OII] gas 
kinematics is decoupled from galaxy rotation in some galaxies and could 
be confined to the nucleus. Two properties of kinematically anomalous 
galaxies have emerged so far: (1) All but one have close companions. 
This suggests that enhanced star formation activity may be the result 
of merger events. (2) Some of these galaxies appear to be of early-type. 
Kinematically ``anomalous'' field galaxies make up 25$\%$ of the field 
sample. This is similar to the fraction of blue-nucleated galaxies 
observed in HST images of galaxies at z $\sim$ 0.6 (Schade {\it et al.} 
1995). 

In one galaxy at z = 0.35, the [OII] line was {\it donut-shaped}. This 
line shape could be interpreted in two ways: either (1) the line is made 
of two rotation curves from two galaxies very close together, or (2) 
the line arises from an expanding supershell, presumably driven by 
supernova winds from a massive starburst. If option (2) is correct, 
then the supershell has a diameter of 2.5 h$^{-1}$ kpc and a rest-
frame expansion velocity of about 80 km/s. These characteristics are 
amazingly similar to those of supershells observed in local dwarf 
irregular galaxies (Marlowe {\it et al.} 1995, Martin 1995). In another 
galaxy at z = 0.42, there is a relatively strong [OII] source unresolved 
both spectrally and spatially superimposed over a fainter rotational 
[O II] component. The source could be a giant HII region 2.6 h$^{-1}$ 
kpc from the center of the galaxy. More details on these unusual sources 
will appear elsewhere (Simard and Pritchet 1996).

\subsection{An Intermediate Redshift Tully-Fisher Relation} \label{tfplot}
Figure~\ref{fig1} shows kinematical evidence for luminosity evolution at 
intermediate redshifts. The locus of the local H$_\alpha$ rotation 
velocity$-$B band Tully-Fisher (TF) relation for all morphological types 
(from the data of Mathewson {\it et al.} 1992) is used as a reference. 
Solid circles are the rest-frame V$_{\rm rot}$ sin $i$ 's versus 
rest-frame B magnitude for the kinematically normal galaxies in our sample 
--- i.e. galaxies with [OII] scale length consistent with their broad-
band scale length. B-band k-corrections were computed using the tables 
of Frei and Gunn (1994). The upper long dashed line is an unweighted 
linear fit to all local morphological types. This linear fit was then 
shifted by $\Delta M_{B_0}$ = $-$1.0 mag (middle dashed line) and 
$\Delta M_{B_0}$ = $-$2.0 mag (lower dashed line) to represent various 
degrees of luminosity evolution.

Galaxies in our sample are $\sim$1.5$-$2.0 magnitudes brighter than 
expected from their rotation velocity, given the local TF relation 
defined by the Mathewson {\it et al.} sample. Such a large local sample 
clearly shows the scatter in the local TF relation. Locally, there are 
no systematic shifts between loci of different morphological types, 
but it is obvious that certain types (e.g. T = 6) display larger 
dispersions.


%
%

\section{Discussion}
The scatter in the local TF relation has a direct impact on the exact 
amount of luminosity evolution derived from our observed TF relation at 
intermediate $z$. This is because scatter in the local TF relation may 
correlate with star formation rates, and hence emission line strengths. 
In order to accurately measure magnitude offsets relative to the local 
Tully-Fisher relation, each galaxy in our CFHT sample (or, for that matter, 
any other sample) should be compared to the local Tully-Fisher relation 
for galaxies with {\it similar} [OII] emission line strengths. This is not 
possible at present. A large, homogeneous sample of local emission-line 
strengths, rotation velocities, morphologies and absolute magnitudes is 
required to settle this issue. This will remain a major limitation of 
internal kinematics studies at high redshifts as long as technical 
requirements restrict such studies to strong emission-line objects.

The luminosity-dependent luminosity evolution scenario neatly reconciles 
the various amounts of luminosity evolution seen in the surface brightness 
and internal kinematics studies. At the low end of the galaxy mass 
spectrum, Koo {\it et al} (1995) looked at compact, unresolved galaxies 
with narrow emission line widths. The line widths of these galaxies were 
2$-$3$\times$ lower than expected from the TF relation of normal spiral 
galaxies, implying large amounts of evolution ($\sim$ 3 mag). These 
low-mass galaxies had linewidths similar to those of local HII galaxies. 
At the high end of the galaxy mass spectrum, the Keck study of Vogt 
{\it et al.} (1996) selected galaxies that were intrinsically large 
(r$_d$ $\geq$ 3.0 kpc) and bright (M$_{B}$ $\leq$ $-$20.7). These 
galaxies were also intrinsically massive, with typical rotation 
velocities of 200 km/s. The increase in B luminosity with respect 
to the local TF relation was less than 0.6 mag.

Our CFHT sample occupies a niche in size and mass {\it between} the 
above two Keck samples. The CFHT galaxies are typically a full magnitude 
fainter than the objects in the Vogt {\it et al.} study. They are also 
intrinsically smaller with typical disk scale lengths less than 2.0 kpc, 
and some of them are barely resolved. They are also less massive, with 
rotation velocities $\sim$ 100 km/s, but they are nonetheless more 
massive than the galaxies observed by Koo {\it et al}.. If mass is taken 
as an indicator of the luminosity that a galaxy would have had in a 
quiescent phase, then all three internal kinematics studies can be 
understood with luminosity-dependent luminosity evolution. 

The B-band surface brightness $\mu_0$(B) of field disk galaxies undergoes 
strong evolution over the redshift range 0.1 $<$ z $<$ 0.6 compared to 
two different local log r$_d$$-$M$_{AB}$(B) relations: the Freeman law 
and an empirical z = 0.06 relation for galaxies in Abell 2256 (Schade 
{\it et al.} 1996). At a redshift of 0.5, $\Delta\mu_0$(B) is equal to 
$-$1.1 mag. This is consistent with or perhaps slightly less than the 
evolution seen in our CFHT sample, but it is certainly more than the 
amount of evolution seen in the Keck sample of Vogt {\it et al.}. Looking 
at Figure 1 of Schade {\it et al.}, there is a hint that surface 
brightness evolution depends on disk scale length: smaller galaxies 
evolve more drastically than large galaxies. The effect is particularly 
noticeable in the highest redshift bin where the log r$_d$$-$M$_{AB}$(B) 
relation clearly curves ``down''. Taking M$_{B}$ $\simeq$ $-$21 and 
$<$r$_d$$>$ = 4.3 kpc (H$_0$ = 75), one can see on the Schade diagram 
for 0.45 $<$ z $<$ 0.65 that a number of bright, large galaxies at 
(log r$_d$  = 0.8, M$_{AB}$(B) = $-$21) show little or no evolution 
-- similar to what is observed in the Keck sample! 

The diversity of [OII] kinematics and the significant amount of luminosity 
evolution seen in our survey illustrate the power of internal kinematics 
as a probe of the nature of intermediate redshift galaxies. Although 
samples are small compared to those of redshift surveys, they already 
provide exciting direct evidence that luminosity-dependent luminosity 
evolution may indeed be at the root of the faint galaxy excess problem. 
Tying internal kinematics to other galaxy properties on a galaxy by 
galaxy basis will play a key role in our understanding of galaxy evolution 
at intermediate redshifts.


\acknowledgments

We thank the Canadian Time Allocation Committee for the CFHT for generous 
grants of observing time, and the CFHT organization for the technical 
support which made this project possible. We also thank the Canadian 
Astronomical Data Center (CADC) for generous allocations of computer 
time. We gratefully acknowledge financial support from the Natural 
Sciences and Engineering Research Council of Canada through an operating 
grant to CJP and a Postgraduate Fellowship to LS.

\clearpage
\begin{deluxetable}{lcccccrr}
\tablecaption{Target galaxies: Characteristics and Results. \label{tbl-1}}
\tablehead{
\colhead{ID} & \colhead{z} & \colhead{Gunn} & \colhead{Gunn} & 
\colhead{W$_{3727}$} & \colhead{M$_{B_0}$} & \colhead{V$_{\rm {rot}}$ sin $i$} 
& \colhead{r$_d$} \\ 
\colhead{}  & \colhead{} & \colhead{r} & \colhead{$(g-r)_{obs}$} & 
\colhead{(\AA)} & \colhead{} & \colhead{(km/s)} & \colhead{(h$^{-1}$ kpc)} \\
\colhead{(1)}  & \colhead{(2)} & \colhead{(3)} & \colhead{(4)} & \colhead{(5)} 
& \colhead{(6)} & \colhead{(7)} & \colhead{(8)}\\
} 
\startdata
A2390-101033\tablenotemark{a} \tablenotemark {c} & 0.2460 & 18.90 & 0.70 & 
55 & $-$20.3 & $-$212 $^{+\hskip 8 true pt 9}_{-\hskip 4 true pt 11}$ & 
0.09 $^{+0.02}_{-0.02}$ \nl
A2390-100686\tablenotemark{n} & 0.3822 & 21.22 & 0.67 & 21 & $-$19.4 & 
91 $_{-\hskip 4 true pt 39}^{+\hskip 4 true pt 21}$ & 2.3 $_{-0.40}^
{+0.50}$ \nl
A2390-350416\tablenotemark{n} & 0.2558 & 20.14 & 0.27 & 35 & $-$19.5 & 
62 $_{-\hskip 4 true pt 21}^{+\hskip 4 true pt 18}$ & 1.9 $_{-0.40}^
{+0.40}$ \nl
A2390-350471\tablenotemark{n} & 0.2559 & 19.40 & 0.28 & 21 & $-$20.3 & 
171 $_{-\hskip 4 true pt 29}^{+\hskip 4 true pt 26}$ & 2.2 $_{-0.50}^
{+0.70}$ \nl
E1512-301037A\tablenotemark{n} & 0.3457 & 21.41 & 0.19 & 41 & $-$19.1 & 
77 $_{-\hskip 4 true pt 24}^{+\hskip 4 true pt 22}$ & 1.8 $_{-0.50}^
{+0.20}$ \nl
E1512-301037B\tablenotemark{n} & 0.3457 & 21.41 & 0.19 & 41 & $-$19.1 & 
86 $_{-\hskip 4 true pt 18}^{+\hskip 8 true pt 8}$ & 0.8 $_{-0.20}^
{+0.40}$ \nl
E1512-101526\tablenotemark{a} & 0.4026 & 20.36 & 1.12 & 48 & $-$20.2 & 
$-$222 $_{-\hskip 4 true pt 33}^{+143}$ & 0.1 $_{-0.10}^{+0.40}$ \nl
E1512-201429\tablenotemark{n} & 0.4231 & 20.44 & 0.56 & 26 & $-$20.4 & 
151 $_{-\hskip 4 true pt 11}^{+\hskip 4 true pt 13}$ & 1.0 $_{-0.10}^
{+0.20}$ \nl
E1621-100515 & 0.3455 & 20.00 & 0.42 & 22 & $-$20.3 & $-$237 $_{-\hskip 
8 true pt 7}^{+\hskip 4 true pt 18}$ & 0.8 $_{-0.10}^{+0.20}$ \nl
A2390-100225 & 0.3829 & 21.65 & 0.89 & 26 & $-$18.9 & 61 $_{-107}^
{+121}$ & 0.6 $_{-0.40}^{+0.50}$ \nl
A2390-101084\tablenotemark{a} \tablenotemark {c} & 0.2302 & 17.31 & 
0.68 & 110 & $-$21.7 & $-$271 $_{-\hskip 4 true pt 12}^{+\hskip 4 
true pt 12}$ & 0.4 $_{-0.06}^{+0.03}$ \nl
A2390-200928\tablenotemark{n} & 0.2645 & 21.51 & 0.22 & 40 & $-$18.2 & 
65 $_{-\hskip 4 true pt 15}^{+\hskip 4 true pt 13}$ & 0.3 $_{-0.10}^
{+0.20}$ \nl
A2390-200802\tablenotemark{n} & 0.3208 & 21.16 & 0.37 & 25 & $-$19.1 
& $-$198 $_{-\hskip 4 true pt 65}^{+\hskip 4 true pt 76}$ & 1.0 $_
{-0.50}^{+0.40}$ \nl
A2390-200372\tablenotemark{n} & 0.3485 & 20.15 & 0.65 & 16 & $-$20.2 
& $-$86 $_{-\hskip 4 true pt 18}^{+\hskip 4 true pt 52}$ & 3.2 $_{-0.60}
^{+0.90}$ \nl
E1512-201845\tablenotemark{a} & 0.3387 & 19.60 & 0.82 & 36 & $-$20.5 & 
385 $_{-\hskip 4 true pt 31}^{+\hskip 4 true pt 22}$ & 0.4 $_{-0.10}^
{+0.10}$ \nl
E1512-201773 & 0.3383 & 20.06 & 0.35 & 26 & $-$20.3 & $-$116 $_{-\hskip 
4 true pt 15}^{+\hskip 4 true pt 13}$ & 0.9 $_{-0.10}^{+0.10}$ \nl
E1512-200730\tablenotemark{a} & 0.4266 & 21.35 & 0.58 & 32 & $-$19.5 
& $-$1 $_{-\hskip 8 true pt 5}^{+\hskip 8 true pt 7}$ & 0.2 $_{-0.10}
^{+0.10}$ \nl
E1512-200334\tablenotemark{n} & 0.4142 & 21.81 & 0.70 & 36 & $-$19.0 
& 60 $_{-\hskip 4 true pt 31}^{+\hskip 4 true pt 46}$ & 1.5 $_{-0.30}
^{+0.30}$ \nl
E1512-200672\tablenotemark{a} & 0.4152 & 21.85 & 0.43 & 29 & $-$19.1 
& 44 $_{-\hskip 4 true pt 91}^{+\hskip 4 true pt 56}$ & 0.3 $_{-0.10}
^{+0.10}$ \nl
E1512-201268\tablenotemark{n} & 0.3412 & 21.77 & 0.39 & 29 & $-$18.6 
& 33 $_{-\hskip 4 true pt 84}^{+\hskip 4 true pt 45}$ & 1.4 $_{-0.20}
^{+0.30}$ \nl
E1512-202096\tablenotemark{n} & 0.4252 & 20.33 & 0.66 & 23 & $-$20.5 
& 28 $_{-\hskip 4 true pt 33}^{+\hskip 4 true pt 39}$ & 2.0 $_{-0.30}
^{+0.20}$ \nl
E1512-201125\tablenotemark{a} & 0.3823 & 21.69 & 0.74 & 15 & $-$18.9 
& 59 $_{-\hskip 4 true pt 92}^{+\hskip 4 true pt 79}$ & 0.5 $_{-0.30}
^{+0.40}$ \nl
\enddata

 
\tablenotetext{a}{Kinematically anomalous galaxy}
\tablenotetext{n}{Kinematically normal galaxy}
\tablenotetext{c}{Serendipitously observed cluster galaxy}

\tablecomments{(1) Galaxy identification code. First part is the CNOC 
cluster name and the second part is the CNOC PPP number; (2) CNOC galaxy 
redshift; (3) CNOC observed Gunn r magnitude; (4) CNOC observed Gunn 
$g-r$ color; (5) Rest-frame [OII] equivalent width. Obtained by dividing 
the observed width by (1+z). Error $\sim$ 10$\%$; (6) k-corrected absolute 
B-band magnitude (H$_0$ = 75 km s$^{-1}$ Mpc$^{-1}$, q$_0$ = 0.5). Error 
= $\pm$0.1 mag; (7) Rest-frame projected [OII] disk rotation velocity 
with 68$\%$ confidence interval; (8) [OII] exponential disk scale 
length with 68$\%$ confidence interval (h = H$_0$/(100 km s$^{-1}$ 
Mpc$^{-1}$), q$_0$ = 0.5). Galaxies with no superscript were rejected 
because cosmic rays hit the [OII] line directly, or simulations showed 
that the [OII] flux was too low to derive reliable kinematical parameters.}
 
\end{deluxetable}

%
%
%

\clearpage




\clearpage

\clearpage



\figcaption[fig1.eps]{Kinematical evidence for luminosity evolution at 
intermediate redshifts. The locus of the local H$_\alpha$ rotation 
velocity$-$B band Tully-Fisher relation for all morphological types as 
defined by data taken from Mathewson {\it et al.} (1992) is used as a 
reference. Solid circles are the V$_{\rm rot}$ sin $i$ 's versus 
rest-frame B magnitude for the kinematically normal galaxies in our sample. 
The upper long dashed line is an unweighted linear fit to all the local 
morphological types. This linear fit was then shifted by $\Delta M_{B_0}$ 
= $-$1.0 mag (middle dashed line) and $\Delta M_{B_0}$ = $-$2.0 mag (lower 
dashed line) to represent various degrees of luminosity evolution. 
\label{fig1}}



\clearpage





\begin{thebibliography}{}

\bibitem[Broadhurst et al. 1988]{broad88} Broadhurst, T. J., Ellis, 
R. S., and Shanks, T. 1988, \mnras, 235, 827
\bibitem[Broadhurst et al. 1992]{broad92} Broadhurst, T. J., Ellis, 
R. S., and Glazebrook, K. 1992, Nature, 355, 55
\bibitem[Colless 1994]{coll94} Colless, M. 1994 in 35$^{th}$ Herstmonceux 
Conference: Wide Field Spectroscopy and the Distant Universe, S. Maddox 
and A. Aragon-Salamanca (eds.), World Scientific Publishing Company, 
Singapore.
\bibitem[Forbes et al. 1995]{forb94} Forbes, D. A., Phillips, A. C., 
Koo, D. C. and Illingworth, G. D. 1996, \apj, 462, 89
\bibitem[Franx 1993]{franx93} Franx, M. 1993, \pasp, 105, 1058
\bibitem[Frei and Gunn 1994]{frei94} Frei, Z. and Gunn, J. E. 1994, 
\aj, 108, 1476
\bibitem[Kennicutt 1992]{kenn92} Kennicutt, R. C. 1992, \apj, 388, 
310
\bibitem[Koo et al. 1995]{koo95} Koo, D. C., Guzman, R., Faber, 
S. M., Illingworth, G. D., Bershady, M. A., Kron, R. G. and Takamiya, 
M. 1995, \apjl, 440, L49
\bibitem[Koo and Kron 1992]{koo92} Koo, D. C. and Kron, R. 1992, 
\araa, 30, 613
\bibitem[Lilly et al. 1995]{lilly95} Lilly, S. J., Tresse, L., Hammer, 
F., Crampton, D. and LeF\`evre, O. 1995, \apj, 455, 108
\bibitem[Lilly 1993]{lilly93} Lilly, S. J. 1993, \apj, 411, 501
\bibitem[Marlowe et al. 1995]{marl95} Marlowe, A. T., Heckman, T. M., 
Wyse, R. F. G. and Schommer, R. 1995, \apj, 438, 563
\bibitem[Martin 1995]{mart95} Martin, C. L. 1995, \baas, 27, no. 4, 1375 
\bibitem[Mathewson et al. 1992]{mathew92} Mathewson, D. S., Ford, V. L. 
and Buchhorn, M. 1992, \apjs, 81, 413
\bibitem[Saha and Williams 1994]{saha94} Saha, P. and Williams, T. B. 
1994, \aj, 107, 1295
\bibitem[Schade et al. 1995]{schade95} Schade, D., Lilly, S. J., 
Crampton, D., Hammer, F., Le F\`evre, O. L. and Tresse, L. 1995, \apjl, 
451, L1
\bibitem[Schade et al. 1996]{schade96} Schade, D., Carlberg, R. G., Yee, 
H. K. C., L\'opez-Cruz, O. and Ellingson, E. 1996, \apjl, Submitted, 
Astro-ph/9604139
\bibitem[Simard and Pritchet 1996]{sim96} Simard, L. and Pritchet, C. 
J. 1996, In preparation
\bibitem[Vogt et al. 1993]{vogt93} Vogt, N. P., Herter, T., Haynes, 
M. and Courteau, S. 1993, \apjl, 415, 95
\bibitem[Vogt et al. 1996]{vogt96} Vogt, N. P., Forbes, D. A., 
Phillips, A. C., Gronwall, C., Faber, S. M., Illingworth, G. D. 
and Koo, D. C. 1996, \apjl, Accepted, Astro-ph/9604096
\bibitem[Yee et al. 1996]{yee96} Yee, H. K. C., Ellingson, E. 
and Carlberg, R. G. 1996, \apjs, 102, 269

\end{thebibliography}
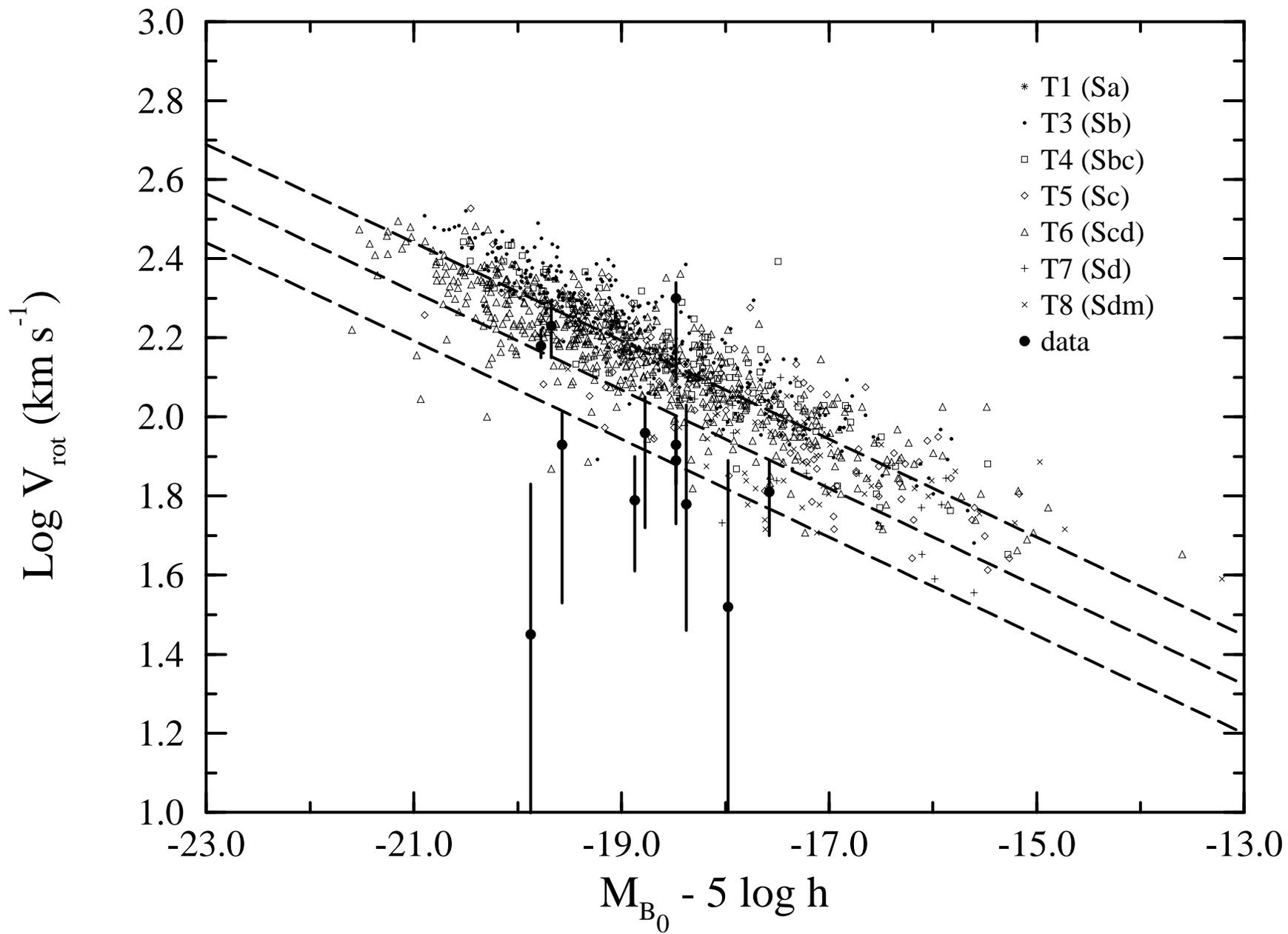
\end{document}